\documentclass[final,5p,times,twocolumn]{elsarticle}

\pdfoutput=1

\usepackage{graphicx}
\usepackage{dcolumn}
\usepackage{bm}
\usepackage{amssymb}
\usepackage{color}
\usepackage{eurosym}
\usepackage{url}
\usepackage[utf8]{inputenc}
\usepackage[T1]{fontenc}
\journal{Applied Surface Science}

\begin{document}
\begin{frontmatter}

\title{Magnetic domains reconfiguration on the Fe$_{3}$O$_{4}$ (110) surface across the Verwey transition by Spin-Polarized Low-Energy Electron Microscopy}

\author[i1,i2]{Clara Gutiérrez-Cuesta}
\affiliation[i1]{Universidad Complutense de Madrid and Unidad Asociada UCM-CSIC(IQF) ''Surface Science and Magnetism of Low-Dimensional Systems'', Madrid 28040, Spain}    
\affiliation[i2]{Instituto de Quimica Fisica Blas Cabrera, CSIC, Madrid 28006, Spain}

\author[i3]{Anna Mandziak}
\affiliation[i3]{Solaris Synchrotron, Czerwone Maki 98, 30-392 Cracow, Poland}

\author[i2]{José Emilio Prieto}
\author[i3,i4]{Paweł Nita}
\affiliation[i4]{Faculty of Physics, Astronomy and Applied Computer Science, Jagiellonian University, ul. prof. Stanisława Lojasiewicza 11, 30-348 Cracow, Poland}
\author[i1]{Arantzazu Mascaraque}
\author[i5]{Usama Choudhry}
\affiliation[i5]{Lawrence Berkeley National Laboratory, Berkeley, USA}
\author[i5]{John Turner}
\author[i5]{Alexander Stibor}
\author[i2]{Juan de la Figuera}

\begin{abstract}
We have studied the (110) surface of Fe$_{3}$O$_{4}$ single crystals by means of spin-polarized low-energy electron microscopy (SPLEEM). After preparation by sputtering and annealing a well defined reconstructed surface was achieved, composed of rows aligned in the [010] direction. By acquiring SPLEEM images along different spin directions the vector magnetization was mapped on the surface, both at room temperature and at a temperature well below the Verwey transition. At room temperature, domains were observed with their magnetization aligned along the two <111> bulk easy axes which are in the (110) surface plane. They presented 180$^{\circ}$, 71$^{\circ}$ and  109$^{\circ}$ Néel-type domain walls.  Below the Verwey transition, the magnetization directions changed to regions where the magnetization was oriented along the in-plane [100] and [001] directions. Those observations can be interpreted as the presence of magnetized regions on the surface where the monoclinic $c$ axis is in-plane in the former, and regions where the $c$ is out-of-plane in an oblique direction in the latter. However, the magnetization was at all times within the surface plane, with no out-of-plane component detected.
\end{abstract}
\end{frontmatter}

\renewcommand{\tablename}{Table}
\section{Introduction}

Magnetite is one of the most remarkable naturally occurring magnetic minerals. References to magnetized “lodestone” (essentially magnetite) appear as early as 600 BCE in Greece and even earlier in Chinese records. As the first magnetic substance known to humankind, it has played a pivotal role in shaping both scientific understanding and technological progress\cite{Blackman1983}. Its combination of high electrical conductivity, chemical stability and elevated Curie temperature\cite{cornell_iron_1996} makes it particularly valuable for magnetic applications. Moreover, paleomagnetic studies rely heavily on its ability to retain information about the Earth’s magnetic field across geological timescales.

At room temperature (RT), magnetite crystallizes in a cubic inverse spinel structure with a lattice parameter of about 8.40~\AA. The oxygen sublattice forms a face-centered cubic array\cite{Parkinson2016SSR}, while Fe cations occupy interstitial tetrahedral (T$_{d}$) and octahedral (O$_{h}$) sites. Fe$^{3+}$ is found in T$_{d}$ sites, whereas O$_{h}$ positions accommodate both Fe$^{2+}$ and Fe$^{3+}$. The ferrimagnetic character of magnetite arises from strong antiferromagnetic coupling between these sublattices. Its Curie temperature is approximately 850 K. It is weakly metallic due to electron hopping between the irons in octahedral sites. Upon cooling, it undergoes a first order phase transition, the Verwey transition\cite{verwey_electronic_1939}, to a insulating phase. This is the oldest known metal-insulator transition and has been subject of much interest since its discovery. Its precise temperature is in the range 115~K to 125~K depending on the sample quality\cite{Iizumi1982}. The low-temperature phase has a monoclinic structure and shows a complex charge order formed by Fe$^{2+}$-Fe$^{3+}$-Fe$^{2+}$ units\cite{senn_charge_2012}. Recent studies have shown that the transition can be optically excited in the nanosecond time range\cite{truc_ultrafast_2024}.
The monoclinic phase can have its $c$ axis pointing along any of the original cubic $\langle 100 \rangle$ axes. Typically, the cubic crystal breaks into a polycrystal with regions with different local orientation of the $c$ axis\cite{kasama2010,juan2013}. The two other monoclinic directions correspond to the perpendicular $\langle 110 \rangle$ cubic directions. In total there are 24 possible different but symmetrically equivalent configurations concerning the relative orientations of the cubic and monoclinic phases\cite{kasama2010}.

The easy axes of bulk magnetite are the $\langle 111 \rangle$ directions at room temperature, as reflected by the negative first-order cubic magnetocrystalline anisotropy constant $K_{c1}$. Upon cooling, the magnetocrystalline anisotropy exhibits a highly non-linear behavior\cite{ReznicekJPc2012}, with an initial increase of the magnitude of $K_{c1}$, up to a maximum at 250~K. Upon further cooling, $K_{c1}$ decreases and changes sign at a temperature typically 10~K higher than the Verwey transition temperature $T_v$; at this temperature the easy axes switch to the $\langle 100 \rangle$ directions, through the so-called spin-reorientation transition. Cooling further, and crossing $T_v$, the crystal becomes monoclinic. In the monoclinic phase, the magnetic easy axis of magnetite is oriented along the c-axis. As shown by Kasama et al.\cite{kasama2010}, extended regions with sizes of tens of micrometers often share an average monoclinic-$c$ axis (the angle between the cubic axis and the monoclinic one is 0.2$^\circ$). Usually those regions present twins along the average monoclinic $c$ direction in the form of lamellas with a width of a fraction of a micron. Within each region with a common $c$ axis, the magnetization is oriented along that axis direction. Different regions have their axis along different cubic axes, and thus different regions present oppositely oriented domains along different cubic [001] directions. We have reported such set of domains at the magnetite (100) surface\cite{LauraPRB2016}. The application of an external magnetic field\cite{Chlan2010,Kosterov2003,Domenicali1950} and/or stress\cite{Coe2012,Kobayashi2009} has been used to select among the possible monoclinic directions through the Verwey transition.

Extensive research has been conducted on the (001) and (111) surfaces of magnetite both for single-crystals and thin films grown on various metallic and oxide substrates\cite{Noh2015, Barbieri1994, Wei2006, MartnGarca2015, delafigueraUltra2013, Maris2006_IEEE, Maris2006_Japanese}. Previous studies utilizing scanning tunneling microscopy (STM) and spectroscopy (STS)
techniques have demonstrated that these surfaces can exhibit multiple terminations depending on the preparation method\cite{Jordan2006, Maris2006, Shimizu2010, Creutzburg2022, Subagyo2007}. By contrast, the (110) surface has received far less attention. Work on this orientation has largely addressed its structural reconstructions and their dependence on preparation conditions. STM and low-energy electron diffraction (LEED) have revealed a characteristic one-dimensional (1$\times$3) reconstruction, which has been linked to a termination involving Fe ions belonging to distinct magnetic sublattices\cite{Jansen1995, Jansen1996, Parkinson2016SS}. Our previous work has reported the magnetic and chemical characterization by X-ray magnetic dichroism in Photoemission Electron Microscopy (XPEEM)\cite{AnnaSciRep2025}.

Although millimeter-sized magnetite crystals lack stable remanence in rocks due to their magnetic softness, their orientation and sectioning along specific crystal planes facilitate investigation. The interpretation of domain structures becomes more straightforward when the exposed surface contains one or more magnetic easy axes, as demonstrated by Chiba in 1983\cite{Chiba1983}. The bulk domain walls belong to three fundamental types, involving changes of 180$^{\circ}$, 109$^{\circ}$ or 71$^{\circ}$ in the orientation of the magnetization inside a domain wall\cite{zdemir1995, Foss1998}. Consequently, a (110) surface, incorporating two different $\langle 111 \rangle$ directions, proves conducive to observe various bulk domain and domain wall configurations without the distortion produced by the shape anisotropy as is the case, for example, of the (100) surface\cite{Halgedahl1987,LauraSciRep2018}. Thus, arrays of these wall types offer representative insights into the internal domain structure when observing the
(110) surface\cite{ozdemir_magnetic_1993}.

Spin-polarized low-energy electron microscopy (SPLEEM) is a technique well suited to determine the orientation of magnetic domains on conductive surfaces\cite{AndreasEPJ2010,BauerBook2014}. It uses the spin-dependent electron reflectivity to image the surface magnetization with nanometer resolution. It is very surface sensitive and requires a highly ordered and clean surface, which often is a drawback when compared with XPEEM. However, it allows for much faster measurements along different magnetization directions. We have previously used it to study the magnetic domains at the surface of magnetite (100)\cite{delafigueraUltra2013} as well as their changes through Verwey\cite{delafigueraPRB2013}, through the spin-reorientation transition\cite{LauraPRB2016} and at high temperature\cite{LauraSciRep2018}. In this work, we present an analysis on the magnetite (110) surface at room temperature and below the Verwey transition by SPLEEM. We have studied its structural and magnetic properties and reconstructed the magnetization vector on surface, at room temperature and below the Verwey transition temperature.

\section{Methods}

The experiments have been performed in the SPLEEM\cite{duden_compact_1995, rougemaille_magnetic_2010} at the Molecular Foundry of the Lawrence Berkeley National Laboratory. This instrument is an Elmitec LEEM III with a spin-polarized electron source based off a Cs-O GaAs cathode emitter, with an approximate spin-polarization of 20\%. Magnetic contrast images correspond to the pixel-by-pixel difference of images acquired with the spin direction along opposite orientations. In such images, the white or black contrast corresponds to the local magnetization oriented along or in opposite orientation of the spin direction used to acquire the first image of the pair.  A spin manipulator allows us to change the spin orientation of the electron beam to any desired direction at the sample. In this way, the magnetization along different directions can be mapped. Combining images with different spin directions it is possible to reconstruct the magnetization vector on the surface. The instrument can also acquire Low-Energy Electron Diffraction (LEED) patterns from selected areas of the surface. The system is equipped with a preparation chamber with an ion gun and an Auger spectrometer.

On the surface of a hat-shaped Fe$_3$O$_4$ (110) crystal from a commercial supplier several square-shaped marks of different sizes were milled by focused gallium ion beam (FIB) to better locate the same region at different temperatures. The single crystal was prepared by a standard procedure of cycles of sputtering with argon at 1000 eV for 10 min and annealing at 600$^\circ$C in an oxygen pressure of 10$^{-6}$ mbar, as has been described in previous works \cite{Jansen1995}, until the Auger spectra did not show any impurities. The crystal is the same that has been used in Ref.~\citenum{AnnaSciRep2025}.

\section{Discussion}

\begin{figure}[htb]
\centering
\includegraphics[width=0.5\textwidth]{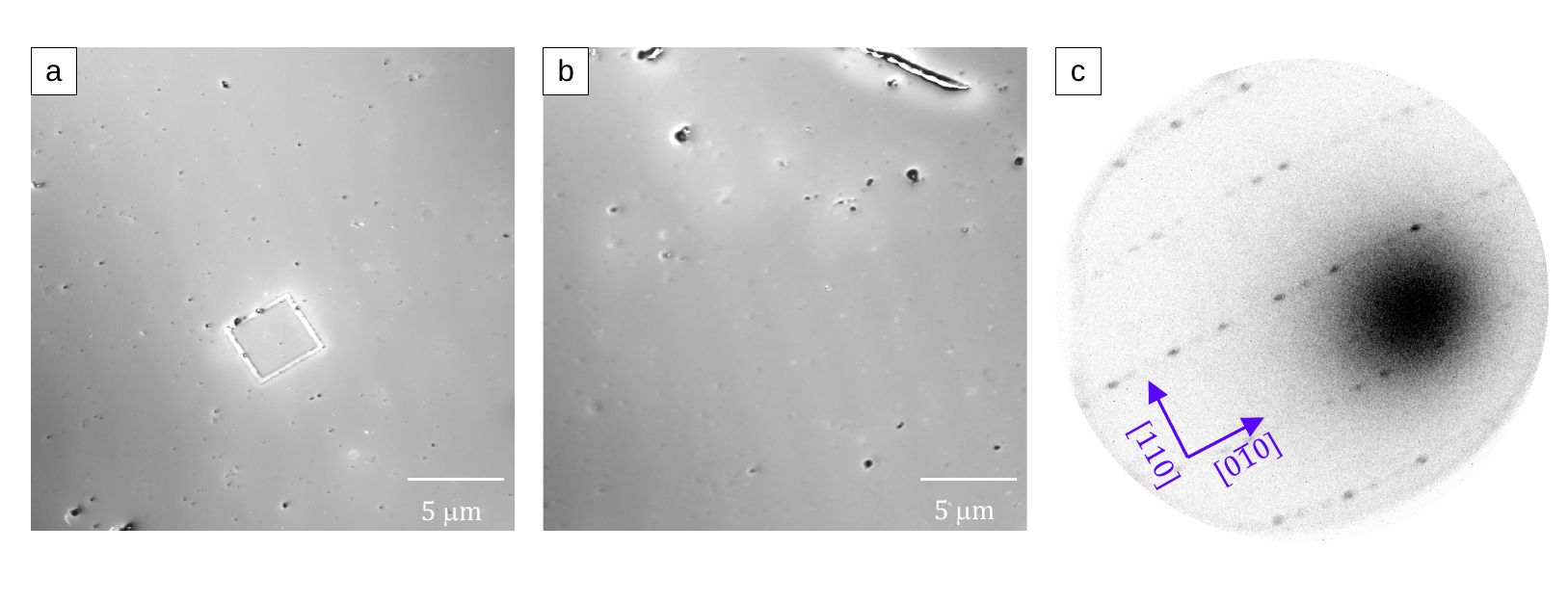}
\caption{a-b) Two representative LEEM images of the clean Fe$_3$O$_4$(110) surface after sputtering and annealing preparation. The square is a feature milled by FIB. c) LEED pattern of the magnetite (110) surface at 45 V from the same area. The orientation of the crystal is shown.}
\label{fg:LEEM}
\end{figure}

After several cycles of mild sputtering and annealing in oxygen, the surface of the magnetite crystal has the aspect shown in Figure~\ref{fg:LEEM}a-b. Small features can be observed, as well as black dots which are presumably due to carbon contamination. The Auger spectra (not shown) from this surface only detected iron and oxygen. The square mark at the center was carved by FIB prior to the SPLEEM experiments. After the preparation procedure, the LEED pattern from the surface is shown in Figure~\ref{fg:LEEM}c, which corresponds to the reconstruction reported previously\cite{Jansen1995,Parkinson2016SS}. It corresponds to a (1$\times$3) surface reconstruction with the 0.3~nm periodicity along the $[1\overline{1}0]$ direction, and a 2.5~nm periodicity along the [001] direction. While the reconstructed structure has not been determined in detail, it has been proposed that it is due to exposure of (111) microfacets\cite{Parkinson2016SS}.

\begin{figure}[htbp]
\centering
\includegraphics[width=0.5\textwidth]{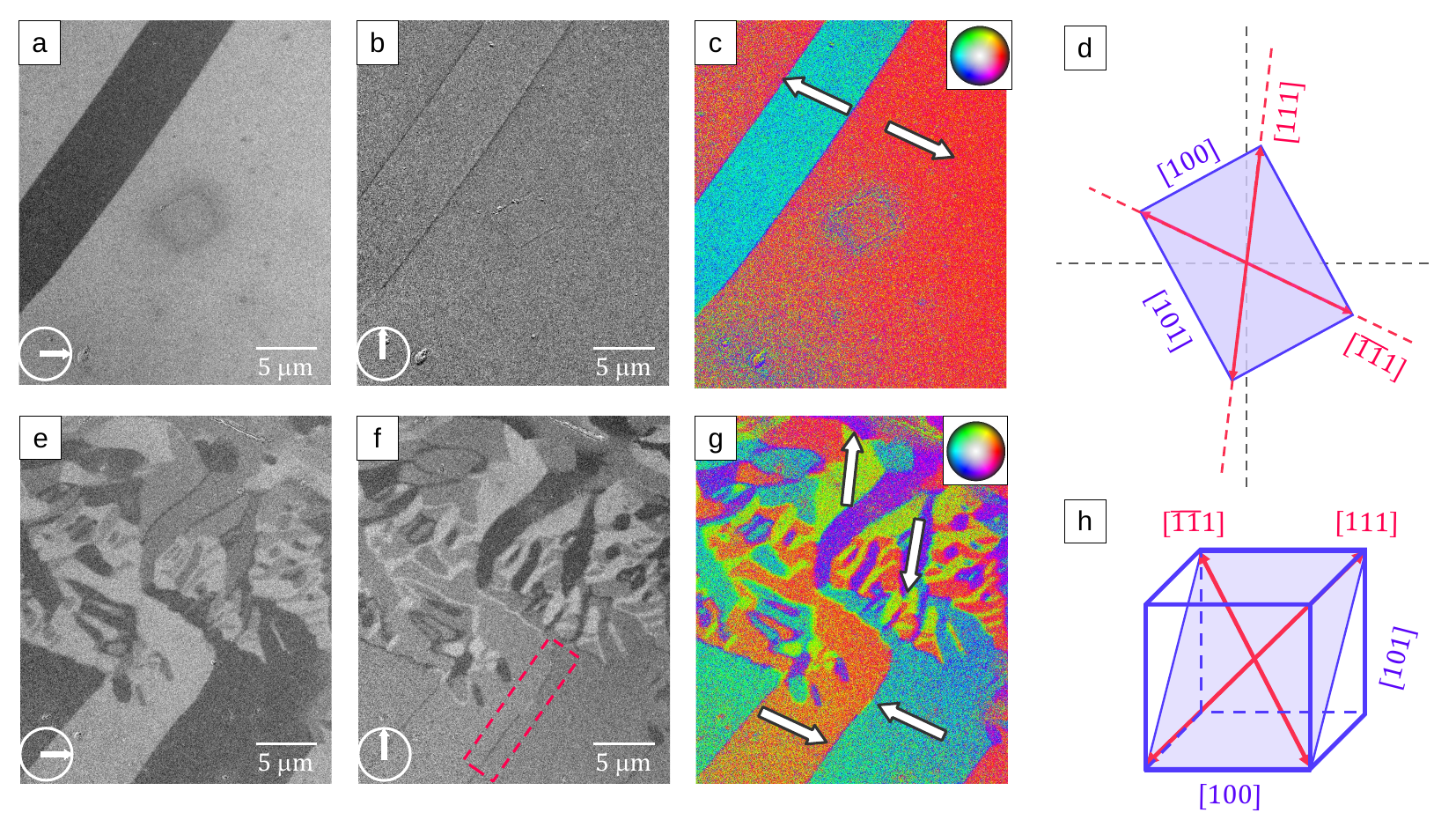}
\caption{a) SPLEEM image of the Fe$_3$O$_4$(110) surface, with the spin-polarization along the x-axis of the image. The blurry region at the centre corresponds to the FIB mark. b) Same, with the spin-polarization along the y-axis of the image. The start voltage in both images is 3.8 V and the FoV, 50$\mu$m. The spin direction is marked with an arrow at the bottom part of each image c) Reconstruction of the vector magnetization from the combination of a) and b) images. The orientation is coded according to the color spin wheel (inset at the right top corner) and it is also marked by arrows in the domains. e) to f) follows the same image sequence: spin polarization along the x-axis, spin polarization along the y-axis and reconstruction of the magnetization vector in a different region of the surface. d) shows a schematic of the orientation of the Fe$_3$O$_4$(110) (unreconstructed) unit cell in purple, the easy magnetization directions corresponding to the <111> in-plane directions in red, and the x/y spin directions in black. h) presents the easy magnetization directions of the (110) surface at room temperature relative to the bulk unit cell.
}        
\label{fg:RT}
\end{figure}

Magnetic contrast images are shown in Figure~\ref{fg:RT}a-c and e-g, recorded on two different regions of the sample. The images in Figure~\ref{fg:RT}a and e correspond to the illuminating electron spin oriented along the x-axis of the images, while Figure~\ref{fg:RT}b and f correspond to the spin direction oriented along the y-axis.  In both groups of images, four contrast levels can be observed: black, white and two shades of gray.
In order to correlate the magnetization vectors with the crystallographic directions, Figure~\ref{fg:RT}d shows the orientation of the (110) surface determined from the LEED pattern in Fig~\ref{fg:LEEM}b. In this scheme it can be observed that the easy [111] and [$\overline{11}$1] are contained in the surface and one of them is close to the y-axis of the images, with only a difference of 6º between the bulk direction and the axis. On the other hand, the [11$\overline{1}$] and $[\overline{11}1]$ make an angle of 26º with the x-axis of the images. This means that the black and white contrast in the Figure~\ref{fg:RT}a,e corresponds to the [$\overline{11}$1] easy-axis, which are mostly aligned with the spin directions, and the gray areas to the [111] one, and the reverse is true for Figure~\ref{fg:RT}b,f. 
Domain walls can also be observed, particularly in Figure~\ref{fg:RT}b, where the domain walls have the same sense along their length. In contrast, in Figure~\ref{fg:RT}f some of the domain walls can be seen to switch sense (from white to black contrast), marked with a red box.
Figures~\ref{fg:RT}c and d show the reconstructed 2D magnetization vector. The color wheel at the right-top of the images indicates the magnetization directions, but to make the interpretation simpler we have also indicated the magnetization directions with arrows in selected places.  The first area, Figures~\ref{fg:RT}a-c with a 7$\mu$m square carved by FIB, contains two large magnetic domains. The reconstruction of the magnetization vector shows that the domains are red and cyan, with long and straight 180º domain walls between them, which are mostly perpendicular to the magnetization domains themselves. The second area, Figures~\ref{fg:RT}e-g, shows a much more complicated domain structure, specially in the upper half. The lower region is quite similar to the previous one, with large domains separated by straight 180º domain walls. But in the upper half, the domain structure consist of much smaller domains with irregular shapes. In this case, we observed four magnetization orientations in domains: cyan, red, yellow and purple, these last two corresponding to the $[11\overline{1}]/[\overline{11}1]$ easy axis direction, again separated by 180º domain walls. Thus implies that we find two oblique easy axes on the surface which the domains follow. The angular separation between the cyan and the yellow domains is 70.53º, and from the cyan to the purple ones is 109.47º, as expected for the Fe$_3$O$_4$(110) surface. Thus the four types of domain walls expected between the in-plane <111> domains are observed. We do not see any evident difference in the surface morphology between both regions. No magnetic contrast along the z-axis spin polarization direction is observed in both areas.

\begin{figure}[htbp]
\centering
\includegraphics[width=0.5\textwidth]{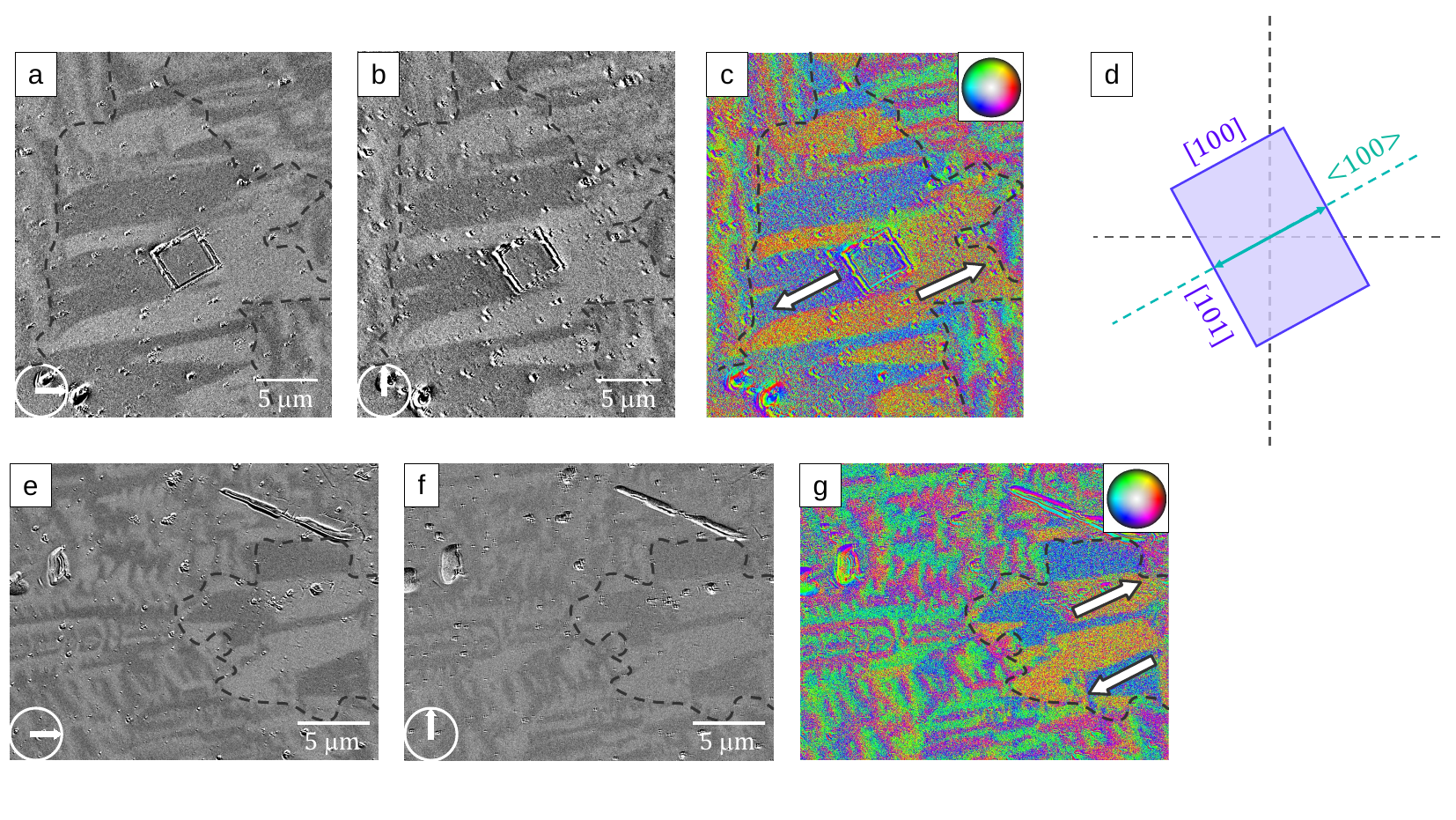}
\caption{a) SPLEEM image of the Fe$_3$O$_4$(110) surface, with the spin-polarization along the x-axis of the image. b) Same, with the spin-polarization along the y-axis of the image. c) Reconstruction of the vector magnetization at the Fe$_3$O$_4$(110) surface, with the orientation coded according to the color spin wheel (inset at the left top corner). d) Schematic of the magnetization directions observed on the (110) surface. e) to g) follow the same sequence as a) to c)} 
\label{fg:LT}
\end{figure}

The sample was then cooled down to 30~K, well below the Verwey transition temperature. In Figure~\ref{fg:LT} we present the domains in the same two areas shown in Figure~\ref{fg:RT}. The domain structure has significantly changed in shape, size and contrast. Figures~\ref{fg:LT}a-b and e-f show that the domains are smaller; they have a streaky aspect. Two regions can be detected: one where the domains are bigger and adjacent domains show the same contrast in both the x- and y-axes spin polarization images, and other where the domains are smaller, more irregular and adjacent domains have opposite contrast in x- and y-axes spin polarization images. If we look at the reconstruction of the magnetization vector in Figures~\ref{fg:LT}c and g, these two different regions are more clearly defined. Areas with the larger domains are yellow and blue (and have mostly the same contrast in the x- and y-spin images), indicating that their magnetization is oriented along the [010]/[0$\overline{1}$0] directions, with  180º domain walls between them. The other region has pink and green smaller domains (the region that shows reversed contrast in the x- and y-spin images). The domains in this area are thus perpendicular to the magnetization in the other areas, and have also 180º domain walls between them: they have domains along [110] and $[\overline{11}0]$ directions.

To understand the domains in the monoclinic phase, we need to consider the possible ways in which the cubic phase gives rise to the low-temperature monoclinic one. A region of the original single cubic crystal can give rise to a monoclinic region with its $c$ axis aligned with any of the bulk <100> cubic directions, i.e. it can have the $c$ axis along either of the [100], [010] and [001] directions. The distribution of domains orientation in the monoclinic polycrystal can be modified by applying a magnetic field or strain while cooling through the Verwey transition. In our present case, no magnetic field was applied, although some strain might arise from mounting the sample in the microscope sample holder.

We first consider the areas in Figure~\ref{fg:LT} that cover domains in blue/yellow, i.e. areas that have their magnetization along the [100]/[$\overline{1}$00] directions. As the monoclinic sub-Verwey magnetite is strongly uniaxial with the easy axis along the monoclinic $c$ axis, this implies that in those areas the monoclinic $c$ axis is along the former cubic [100] direction. In this way the magnetization lies within the surface plane, and there are two different possible magnetic domains. This is the orientation we expect to be favored energetically as both shape and magnetocrystalline anisotropies push the magnetization to be within the surface plane.

\begin{figure}[htbp]
\centering
\includegraphics[width=0.5\textwidth]{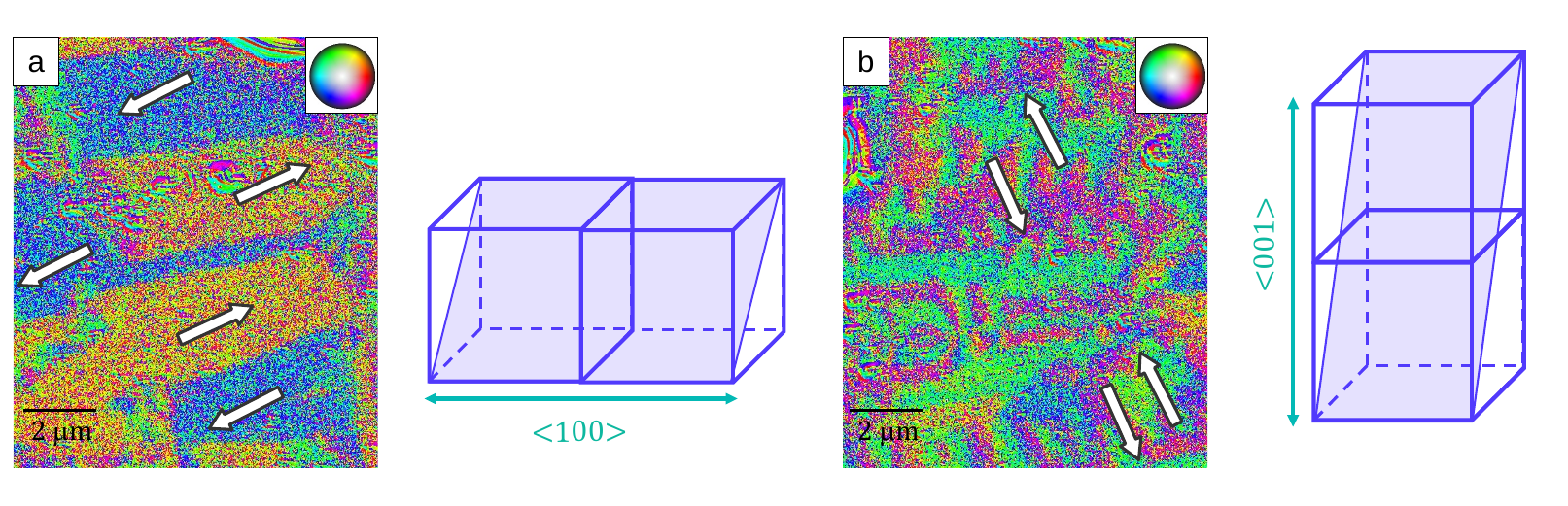}
\caption{a) Area of the magnetization vector reconstruction with big domains that corresponds to the $c$-monoclinic in plane easy axis shown in the structure. b) Other region of the same area with small domains, with the $c$-monoclinic easy axis out of plane. The directions of the magnetization point perpendicular to the direction expected below the Verwey transition.}
\label{fg:monoclinic}
\end{figure}

However, we find other regions where the domains are oriented along the [110] and $[\overline{11}0]$ directions. Again, the domains are oriented in two opposite directions, as expected from an easy axis, with 180º domain walls between them, but that direction does not correspond to a bulk <100> direction. Instead we suggest that in those areas, with pink and green domains, the monoclinic $c$ axis is actually at an oblique angle to the surface, so the projection of the $c$ axis on the (110) surface is along the [110] direction. Figure~\ref{fg:monoclinic} shows a magnification of the area shown in Figure~\ref{fg:LT} with these two regions. As it can be seen, Figure~\ref{fg:monoclinic}a shows the region where the domains are bigger and are aligned with the [100] crystallographic direction, while Figure~\ref{fg:monoclinic}b, with small domains, would correspond to the [001] bulk direction of magnetite. This would imply that in those areas there is a competition between shape anisotropy and magnetocrystalline anisotropy, likely giving rise to smaller domains near the surface. However, we did not observe any magnetic contrast in the z-axis spin direction.

Thus our conclusion is that the two regions observed below the Verwey transition correspond to different orientations of the monoclinic $c$ axis, one within the surface plane and another at an oblique angle. The magnetic domains with the monoclinic axis in the surface plane are not expected to have any out-of-plane magnetization, as both the magnetocrystalline and the shape anisotropy both induce in-plane magnetization. The case of a monoclinic axis at an oblique angle has a competition between magnetocrystalline anisotropy, which is along the monoclinic axis, and the shape anisotropy, wich wants to drive the magnetization in plane. We have not observed an out-of-plane magnetization component in neither area. This is similar to the observation of in-plane magnetization along the [110] directions on the (100) surface\cite{delafigueraUltra2013,LauraSciRep2018}, albeit the magnetocrystalline anisotropy below the Verwey transition is higher that the one of the cubic phase at room temperature. We remark that different monoclinic axis in magnetite below Verwey is a common occurrence in macroscopic crystals and have been observed by transmission electron microscopy by by Kasama et al.\cite{kasama2010}. In fact, this polycrystalline character often complicates structural analysis, and the best bulk determination was thus performed on a micrometric crystal\cite{senn_charge_2012}. Regarding sub-Verwey magnetic domain observations of magnetite crystals, we have reported observations on the (100) surface regions where the monoclinic axis is oriented with the two possible orientations within the surface plane\cite{LauraPRB2016}. We note that even in that case we have unpublished images that suggest that regions with the out-of-plane monoclinic $c$ axis also are present on the surface. We look forward towards studies on the polycrystalline distribution of monoclinic orientations on different magnetite close-packed surfaces.
Another feature reported in many magnetite studies, both in bulk\cite{kasama2010} and at the (100) surface \cite{delafigueraPRB2013} is the observation of monoclinic twins that have a similar monoclinic axis. This produces a "roof"-like surface due to the small misalignment of the uniaxial axis between the twins (about 0.4º), and has been observed in other ferroelastic materials. We have not observed such expected twin domains, but their detection in LEEM requires precise alignment of the electron beam. We thus suggest that those twin domains might be present in the surface that we present below the Verwey transition, and we plan to detect them in further experiments.

\section{Conclusions}
The (110) surface of a Fe$_3$O$_4$ single crystal has been studied by SPLEEM. The surface presented the (1$\times$3) reconstruction observed by LEED. In addition, we have imaged the surface by LEEM. We have associated the directions of the magnetization with the structural directions determined from the LEED pattern. Using the microscope in SPLEEM mode, we have mapped the domains on the surface. At room temperature the domains are oriented along the in-plane <111> directions present in the (110) surface, forming 180º, 109.5º and 70.5º domain walls. In contrast, at the temperature of 30 K well below the Verwey transition, we have detected tens of micrometer sized regions of the surface where the magnetization is either along the [010]/[0$\overline{1}$0] directions, or the [110]/$[\overline{11}0]$ ones. In each region, only 180º domain walls are observed, indicating an easy axis in agreement with the monoclinic character of magnetite below Verwey. We interpret the two regions as areas where the monoclinic $c$ axis is either within the plane, [100], or at an oblique angle to the surface, so that its projection on the surface is along [110].

\section{Acknowledgements}
This research was funded by the SciMat Priority Research Area budget under the program Excellence Initiative - Research University at the Jagiellonian University in Kraków - Grant No. 75.1920.2021 and by Project No. PID2021-124585NB-C31 funded by MCIN/AEI/10.13039/501100011033 and by “ERDF A way of making Europe” and by Mag4TIC(TEC-380) from Comunidad de Madrid (Spain). CGC thanks Universidad Complutense de Madrid and Banco Santander (CT22/25) for financial support. Work at the Molecular Foundry was supported by the Office of Science, Office of Basic Energy Sciences, of the U.S. Department of Energy under Contract No. DEAC02-05CH11231.

\section*{Author Contributions}

Clara Gutiérrez-Cuesta: Formal analysis, Investigation, Writing - Original Draft; Anna Mandziak: Writing - Review and Editing; José Emilio Prieto: Writing - Review and Editing, Formal analysis; Paweł Nita: Writing - Review and Editing; Arantzazu Mascaraque: Writing - Review and Editing, Formal analysis; Usama Choudry: Writing - Review and Editing, Investigation; John Turner: Writing - Review and Editing, Investigation; Alexander Stibor: Investigation, Writing - Review and Editing; Juan de la Figuera: Formal analysis, Writing - Original Draft, Project administration.

\section*{Conflicts of interest}
There are no conflicts to declare.

\section*{Data Avalability}
Data will be available open access at the DIGITAL.CSIC repository with DOI upon acceptance.

\bibliographystyle{elsarticle-num-names} 
\bibliography{magnetite}

\end{document}